\documentclass[%
reprint,
nofootinbib,
 amsmath,amssymb,
 aps,
prl
]{revtex4-1}

\usepackage{graphicx}
\usepackage{siunitx}
\DeclareSIUnit{\pixel}{pixel}

\usepackage{amsmath}
\usepackage{epstopdf, epsfig}
\usepackage{upgreek}
\usepackage[caption=false]{subfig}
\usepackage{array}
\usepackage{microtype}

\usepackage[usenames,dvipsnames]{xcolor}
\usepackage{booktabs}
\usepackage{textcomp} 
\usepackage{xspace} 
\usepackage{tabularx} 
\usepackage{gensymb} 
\usepackage{color}
\usepackage{tikz}

\usetikzlibrary{shapes}

\newcommand{\dashedblackline}{\raisebox{2pt}{\tikz{\draw[-,black,dashed,line width=0.9pt](0,0) -- (5mm,0);}}}

\begin{document}

\title{Initial solidification dynamics of spreading droplets}


\author{Robin B.J. Koldeweij$^{1,2}$}
\email[corresponding author ] {robin.koldeweij@tno.nl}
\author{Pallav Kant$^{1}$}
\author{Kirsten Harth$^{1,3}$}
\author{Rielle de Ruiter$^{4}$}
\author{Hanneke Gelderblom$^{5}$}
\author{Jacco H. Snoeijer$^{1}$}
\author{Detlef Lohse$^{1,6}$}
\author{Michiel A.J. van Limbeek$^{1,6}$}
\affiliation{~1. Physics of Fluids Group \& Max Planck Center Twente for Complex Fluid Dynamics, Department of Science and Technology, J. M. Burgers Center for Fluid Dynamics, University of Twente, 7500 AE Enschede, The Netherlands}
\affiliation{~2. Nano-Instrumentation, TNO, 5612 AP Eindhoven, The Netherlands}
\affiliation{~3. Institute for Physics, Otto von Guericke University Magdeburg, 39106 Magdeburg, Germany }
\affiliation{~4. ASML, 5503 LA Veldhoven, The Netherlands }
\affiliation{~5. Department of Applied Physics, Eindhoven University of Technology, 5600 MB Eindhoven, Netherlands}
\affiliation{~6. Max Planck Institute for Dynamics and Self-Organization, 37077 G\"{o}ttingen, Germany}

\date{\today}

\begin{abstract}
When a droplet is brought in contact with an undercooled surface, it wets the substrate and solidifies at the same time. The interplay between the phase transition effects and the contact-line motion, leading to its arrest, remains poorly understood. Here we reveal the early solidification patterns and dynamics of spreading hexadecane droplets. Total internal reflection (TIR) imaging is employed to temporally and spatially resolve the early solidification behaviour. With this, we determine the conditions leading to the contact-line arrest. We quantify the overall nucleation behaviour, \textit{i.e.} the nucleation rate and the crystal growth speed, and show its sensitivity to the applied undercooling of the substrate. By combining the Johnson-Mehl-Avrami-Kolmogorov nucleation theory and scaling relations for the spreading, we can calculate the temporal evolution of the solid area fraction, which is in good agreement with our observations. We also show that for strong enough undercooling it is the rapid growth of the crystals which determines the eventual arrest of the spreading contact line. 
\end{abstract}


\maketitle

The spreading of a droplet on an undercooled surface is a very complex phenomenon as it instigates several competing physical processes simultaneously: interfacial deformation, contact-line motion and the associated fluid movement, heat exchange between the  droplet and the substrate, and nucleation and growth of a solidified phase within the droplet.
Understanding this process is crucial for a broad range of applications that range from ice accretion on roads \cite{Chen2018}, aircraft \cite{Cebeci2002} and powerlines \cite{Farzaneh2008}, to processes such as soldering \cite{Hayes1999, Attinger2002}, thermal spray coating \cite{Chandra2009} and additive manufacturing \cite{Vaezi2013, Visser2015a}. So far, several investigations have addressed and characterized the intriguing macroscopic behaviour of sessile and impacting droplets on undercooled surfaces. For instance, formation of conical tips during the bulk freezing of a sessile droplet \cite{Enriquez2012} and freezing kinetics along with the final splat morphology of impacting droplets have been investigated in detail \cite{Pasandideh-Fard2002a, Fedorchenko2007, Kong2015, ghabache2016frozen, thievenaz2019solidification, Gielen2019, Kant2020}. Furthermore, nucleation has been studied using top view imaging, by applying a thermal gradient to the atmosphere \cite{Gurganus2011,Tropea2017} or to the substrate \cite{Gurganus2013}. However,  nucleation and growth of crystals at the droplet-substrate interface and its subsequent influence on the droplet spreading has received only little attention \cite{Schiaffino1997, Tavakoli2014,Ruiter2017}. We focus on the interplay between the various phase-transition effects and how they eventually lead to the contact-line arrest.

The arrest of a contact line on an undercooled substrate determines the size and overall shape of the final footprint between the frozen droplet and the substrate. However, due to a lack of direct visualization of the early solidification during droplet spreading, the exact mechanism responsible for contact-line arrest remains debated. So far, various experimental investigations have led to the development of the following explanations: (1) the droplet stops spreading as soon as the contact angle of the spreading liquid reaches the angle of a growing solid front \cite{Schiaffino1997}, (2)  the contact line continues to move until a critical volume solidifies in its vicinity \cite{Tavakoli2014}, (3) the advancing motion of the droplet lasts until the local temperature falls below a threshold at which the crystal growth speed in the vicinity of the contact line becomes equal to the contact-line velocity \cite{Ruiter2017}. 

In this Letter we reveal the sequence of events leading to the contact-line arrest using Total Internal Reflection (TIR) imaging. TIR imaging enables us to characterize the influence of substrate undercooling on the crystal nucleation kinetics as well as the tangential crystal growth along the temporally evolving wetted area. Based on these observations we propose a modelling framework that captures the freezing of an evolving droplet footprint, by combining classical nucleation theory and droplet spreading dynamics.

\begin{figure}
\centering
\includegraphics[width=1\linewidth,clip]{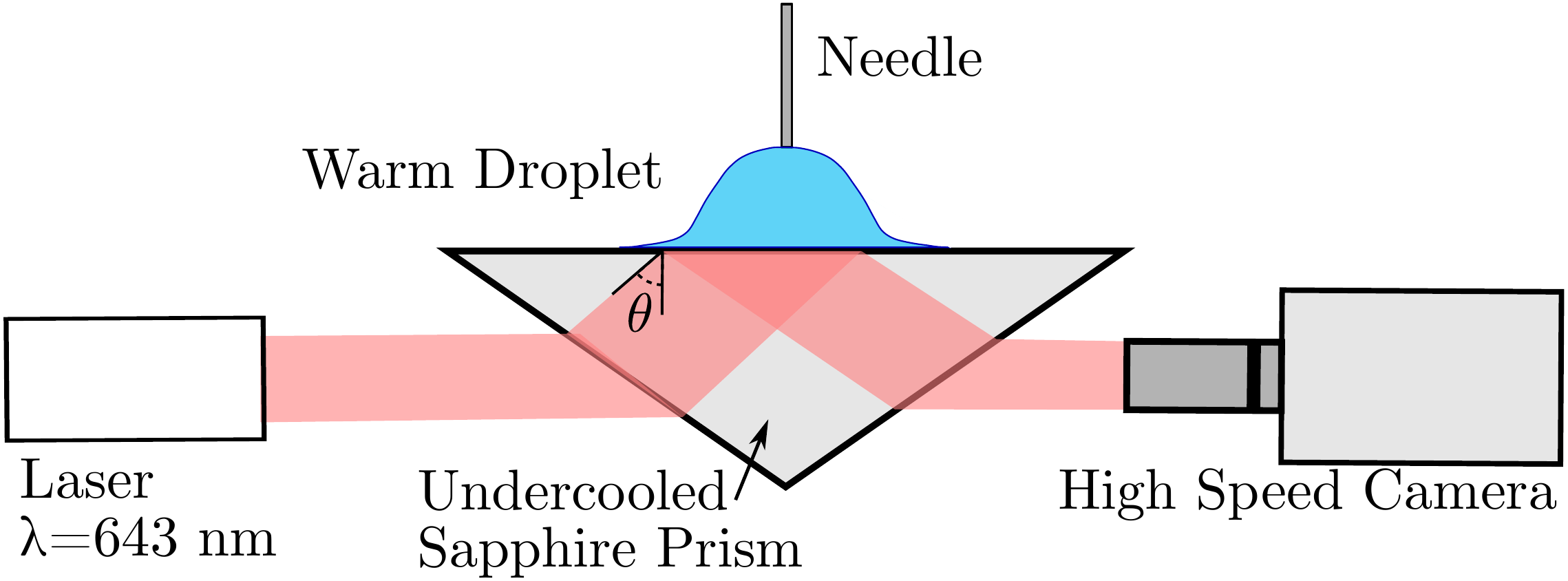}
\caption{Schematic of the experimental setup for total internal reflection imaging.}
\label{fig:setup}
\end{figure}

A schematic diagram of the experimental setup is depicted in figure~\ref{fig:setup}. In a typical experiment, we inflate a droplet of hexadecane to a fixed volume, at the tip of a needle. Hexadecane has a melting point of $T_f=\SI{18}{\degreeCelsius}$. The droplet, of radius $R_0=0.85 \pm 0.05$ \si{\milli\meter}, is then gently lowered (with negligible approach velocity $U$) to the horizontal surface of the undercooled sapphire prism, with temperature $T_s<T_f$. Upon contact, the droplet spreading and freezing is recorded in bottom view via TIR using a high-speed camera connected to a long-distance microscope at $30000$ frames per second. Note that, in contrast to the previously described TIR setups that can measure nanometric thin air-films beneath impacting droplets \cite{Kim2007,Kolinski2012,Khavari2015,Shirota2017}, our setup allows for direct visualization of the solidified phase. This is achieved by choosing the angle $\theta$ of incidence of the laser ($\lambda = 634$ nm) such that total internal reflection occurs not only at the sapphire-air interface but also at the sapphire-hexadecane interface. The solidified material can be visualized owing to the localized scattering of the evanescent wave by the solid particles. Furthermore, the position of the contact line is clearly visible in the images, due to the sudden jump in refractive index between air and hexadecane. A similar setup was recently used in Ref.~\cite{Kant2020}. Details about the visualization, the experimental setup and the material properties can be found in the Supplementary Materials.

\begin{figure}
 \centering
\includegraphics[width=1\linewidth,clip]{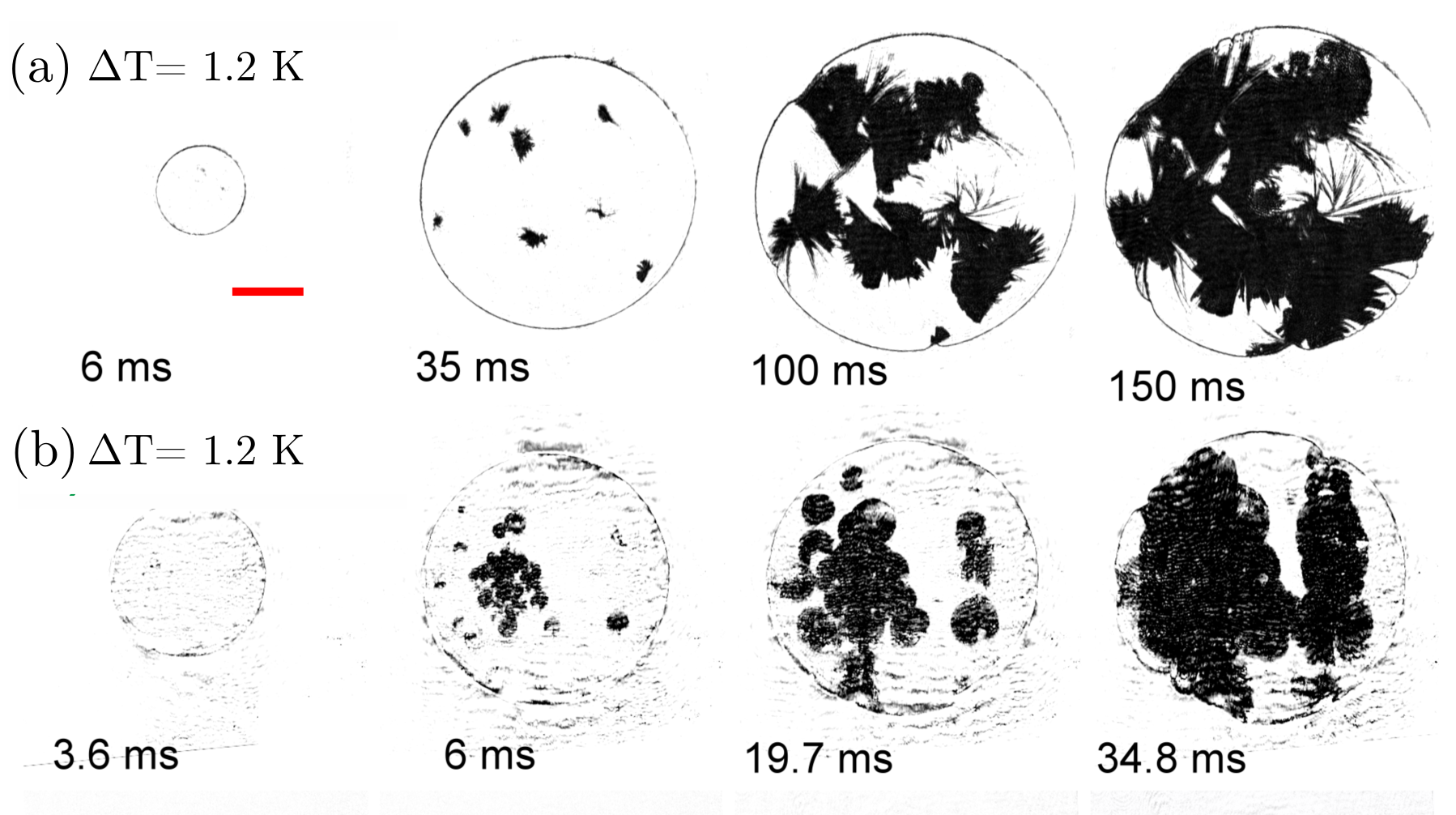}
\caption{Characteristic sequences of hexadecane drops spreading on a sapphire prism of varying temperature, the red bar indicates a length of 1 mm: (a) $\Delta T$=\SI{1.2}{\kelvin}: Random nucleation with subsequent dendritic growth. (b) $\Delta T$= \SI{2}{\kelvin}: Continuous nucleation with subsequent radial crystal growth. Note the different time scales in (a) and (b).}
\label{fig:spottypes}
\end{figure}

Different types of solidification behaviour are observed, for different undercooling $\Delta T = T_f - T_c$. Here, $T_c$ is the droplet-substrate contact temperature, approximated as $T_s+\left(T_d-T_s\right)/\left(1+e_s/e_d\right)$ \cite{Carslaw1959}, with $e=\sqrt{k \rho c_p}$ the thermal effusivity, and the subscripts $s$ and $d$ denoting the substrate and the droplet, respectively. 
Sequences of snapshots in figure~\ref{fig:spottypes} show the freezing behaviours of droplets spreading on substrates at different $\Delta T$. For low undercooling (figure~\ref{fig:spottypes}(a)), nucleation initially occurs only at a few locations that are randomly distributed over the droplet footprint. Subsequently, these crystals nuclei grow into needle-shaped structures: columnar dendrites. The nucleation rate and the morphology of the growing crystals change significantly for a slight increase in undercooling. At higher $\Delta T$ (figure~\ref{fig:spottypes}(b)), a considerable increase in the amount of crystals is observed. Interestingly, in this case, the enhanced nucleation rate is followed by axisymmetric growth of crystal nuclei, seen as seemingly circular footprints (figure~\ref{fig:spottypes}(b)). However, a close inspection reveals that these are still constituted of dendritic patterns. Note that in he phase-transition effects only initiate after a lag time $\tau_g$ \cite{sohnel1988interpretation}. In our experiments,  $\tau_g$ varies from a few microseconds to a few seconds, respectively, at the largest and smallest of the $\Delta T$ employed in our experiments. 
The increase in the nucleation rate at higher $\Delta T$ is directly related to the corresponding decrease in the activation energy for liquid-solid transformation. For the creation of a solid nucleus, this can be considered as the sum of the surface energy between the newly created particle and the bulk, and the released latent energy in the transformed volume of this small nucleus. For a nucleus growing on a surface (heterogeneous nucleation), this critical energy can be approximated as $E_a=\left({16\pi}/{3} \right) {\gamma_{ls}^3}f\left(\theta_{ls}\right) / \left(\Delta g\right)^2 $ \cite{davis2001theory}, with a geometrical correction factor $f\left(\theta_{ls}\right)$, that depends on the contact angle $\theta_{ls}$ of a crystalline deposit with the foreign solid surface \cite{Mullin2001}. Here $\gamma_{ls}$ is the interfacial tension between the liquid and solid hexadecane and $\Delta g=\Delta S_{fus} \Delta T$ is the free energy difference between the liquid and solid phase, with the entropy of fusion $ \Delta S_{fus}=6.28 \cdot 10^5 \si{\joule\per\meter\cubed\per\kelvin}$ \cite{Herhold1999}.
Note that $E_a$ varies as $\left(\Delta T\right)^{-2}$. Accordingly, in our experiments for $\Delta T < 1 K$, we do not observe any nucleation at the experimental timescale ($\sim 5$ sec). Conversely, the droplet footprint instantly solidifies upon touching the substrate for $\Delta T > 2.9 \si{\kelvin}$, when the $E_a$ decreases with 90\%. \\

\begin{figure}
 \centering
\includegraphics[width=1\linewidth,clip]{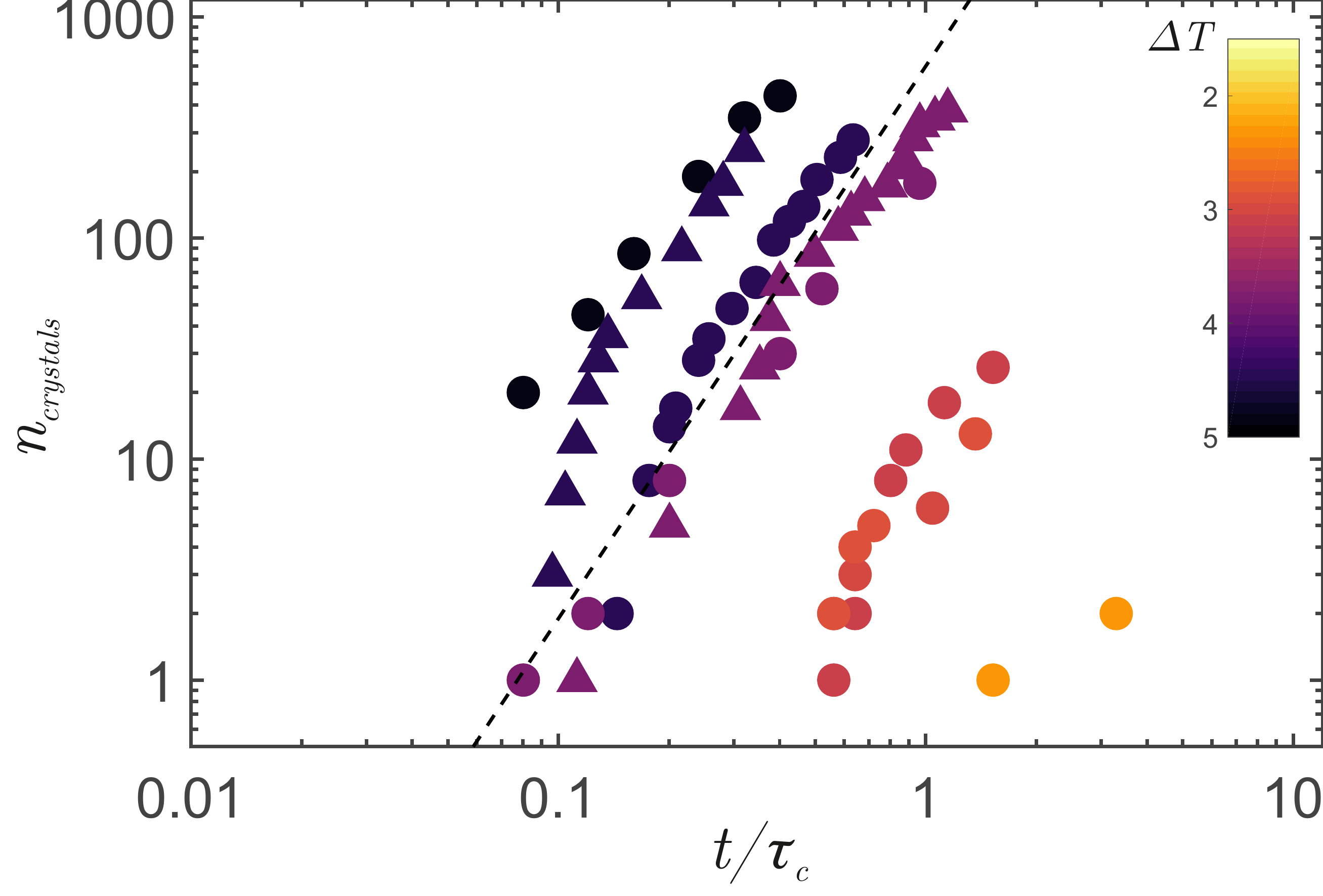}
\caption{Number of distinct crystals as function of dimensionless time. Separate experiments are denoted with different symbols. The dashed line (\protect\dashedblackline) indicates the scaling $ \propto \left(t/{\tau_c}\right)^\frac{5}{2}$  (equation~(\ref{eq:nucleationfull})). The capillary timescale $\tau_c=\left(\rho R_0^3 /\sigma\right)^{1/2}$. The colorbar shows the temperature difference $\Delta T=T_f - T_c$.}
\label{fig:nucleation}
\end{figure}
To quantify the nucleation kinetics at the early-times of droplet spreading, we measure the total number of growing crystals $n_\mathrm{crystals}$ as a function of time. The result is shown in figure~\ref{fig:nucleation}, where time is rescaled by the capillary time $\tau_c=\left(\rho R_0^3 /\sigma\right)^{1/2}$, with the surface tension $\sigma$ between the liquid and air. For significantly large undercooling $\Delta T$, it follows a power-law behaviour, that we will argue to be $n_\mathrm{crystals} \propto \left(t/{\tau_c}\right)^\frac{5}{2}$.

To rationalise this behaviour, we employ classical nucleation theory.  The nucleation rate per unit volume is estimated as $J_0 = A  \exp\left({\frac{-E_a\left(f\left(\theta_{ls}\right)\right)}{k_B T_{f}}}\right)$, where $k_B$ is the Boltzmann constant, and $A$ the attempt frequency per unit volume. The amount of crystals $n_\mathrm{crystals} (t/\tau_c)$ is then obtained by multiplying $J_0$ by the available volume for nucleation, which is estimated by the wetted area $\pi \left(R(t)\right)^2$ times the thermal penetration depth $\delta_{th} \propto \sqrt{\kappa t}$, with $\kappa=k/\left(\rho c_p\right)$ the thermal diffusivity of the liquid. The number of growing crystals is then described by:
\begin{equation}\label{eq:nucleationfull}
n_{\rm{crystals}} \propto J_{\rm{0}} \int_0^{t/\tau_c} R^2 \delta_{\rm{th}} \mathrm{d}t.
\end{equation}
In our experiments, the wetting dynamics of a droplet is indistinguishable from the iso-thermal spreading, as found in Refs.~\cite{Biance2004,Bird2008,Winkels2012,wildeman2016spreading}. The wetting follows the spreading law  $R/R_0 \propto \left(t/\tau_c\right)^{1/2}$ until the contact line suddenly stops advancing due to the solidification, see Supplementary Materials. To close the problem, we assume that the early-time spreading dynamics of the droplet remains unaffected by the nucleation. Combining this spreading law with equation \eqref{eq:nucleationfull}, we derive $n_{crystals} \propto \left(t/\tau_c\right)^{5/2}$, which is consistent with the experimental data (figure~\ref{fig:nucleation}). Note that for very small $\Delta T$ (orange circles in figure~\ref{fig:nucleation}) the amount of crystals does not follow this power law. Due to the low surface energy, $\tau_g$ may even be larger than $\tau_c$. Surface impurities can lead to random nucleation in this case.

\begin{figure}
 \centering
\includegraphics[width=1.\linewidth,clip]{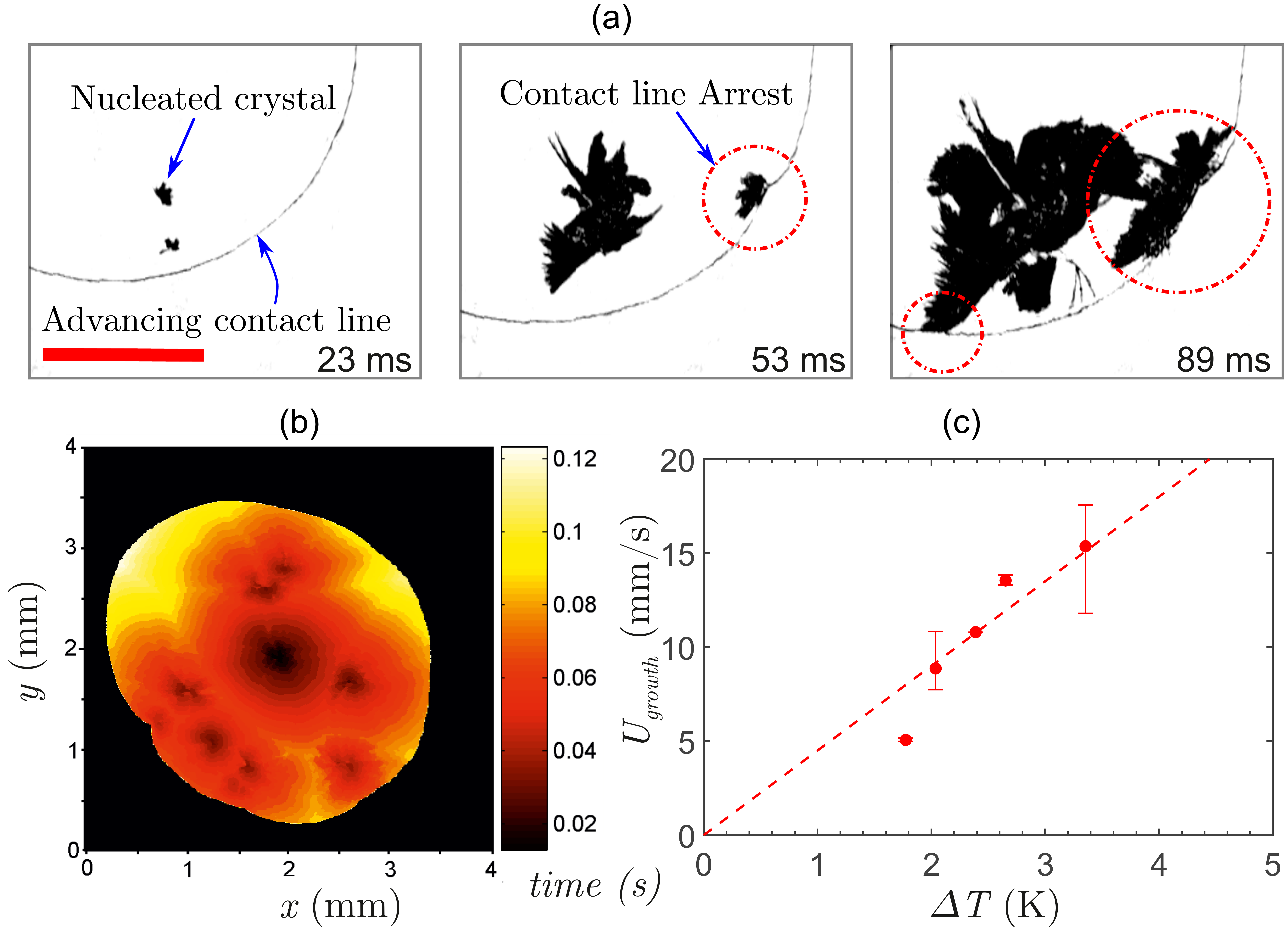}
\caption{
Contact line arrest and crystal growth: (a) Nucleation near the contact line, leading to local contact-line arrest (red circles), the red bar indicates a length of 1 mm, $\Delta T=\SI{1.0}{\kelvin} $. (Left panel) Nucleation far away from the contact line. (Center panel) Nucleation close to the contact line leads to contact-line arrest. (Right Panel) Crystals growing towards the contact line lead to contact line arrest. (b) Footprint of a solidified droplet with the temporal evolution (color variation) of solidification during drop spreading starting from sequentially formed isolated crystals, $\Delta T= \SI{3.4}{\kelvin}$. (c) Growth velocity as a function of the undercooling $\Delta T$. The errorbars indicate the minimal and maximal growth velocity at a certain temperature. The dashed line shows the growth velocity $U_{\rm g}=\beta \Delta T$, with $\beta \approx \SI{4.5}{\milli\meter\per\second\per\kelvin}$. }
\label{fig:pinning}
\label{fig:growth}
\end{figure}

We are now in a position to identify the mechanism that leads to the sudden arrest of the moving contact line. Our experiments reveal that the local nature of the interactions between growing crystals and the moving contact line is one reason that leads to its arrest. Figure~\ref{fig:growth}(a) highlights two distinct events that exemplify the physical mechanism responsible for this contact-line arrest.  In the case that a crystal nucleates at the contact line (figure~\ref{fig:growth}(a), centre panel), it immediately arrests the advancing motion locally. This random event at the moving contact line is caused by local heterogeneities on the substrate. In contrast, a crystal nucleating far away from the contact line (figure~\ref{fig:growth}(a), left panel) does not affect its motion immediately, but if its growth catches up with the advancing contact line, it locally arrests the spreading (figure~\ref{fig:growth}(a), right panel). Consequently, in both cases the droplet footprint evolves non-axi-symmetrically.\\

The motion of the contact line was hypothesised to arrest when the advancing velocity of the contact line becomes equal to the crystal growth speed for the applied under-cooling. This hypothesis leads to a scaling law for the arrest radius $R_{end}$~\cite{Ruiter2017}: 

\begin{equation}\label{eq:clinearrest}
R_{end}/R_0 \propto R_0/\left( \tau_c U_{g} \right),
\end{equation}
with $U_{g}$ the crystal growth speed, which was approximated as $U_{g} = \beta \Delta T$ \cite{davis2001theory}, where the kinetic undercooling coefficient $\beta$ (with units \si{\meter\per\second\per\kelvin}) is a fitting parameter. Here, we directly obtain the crystal growth speed, and with that the undercooling coefficient $\beta$, by measuring the temporal growth of several crystals after their nucleation at the droplet-substrate contact area for different substrate undercooling. Note that $T_c$ is the natural choice instead of the substrate temperature $T_s$ for this undercooling, as the solidification is dictated by the contact temperature. Figure~\ref{fig:growth}b shows a typical evolution of crystals after nucleation in the wetted area. The crystal growth speed shows a linear dependence on $\Delta T$ (figure~\ref{fig:growth}c), resulting in $\beta \approx \SI{4.5}{\milli\meter\per\second\per\kelvin}$ This value is slightly smaller than that for hexadecane spreading on copper obtained from a fit of the data in Ref.~\cite{Ruiter2017}. With this independently measured kinetic cooling coefficient, we can now directly test the prediction for contact line arrest, equation (2), by comparing to our measurements. A very good agreement is found, using a prefactor of $0.18$, with a relative error within 20\% for most data (see Supplementary Materials). However, our experimental results significantly deviate from the model at low undercooling $\Delta T < \SI{3}{\kelvin}$. We believe that the implicit assumptions (infinitely small lag time $\tau_g \sim 0$ and high probability of nucleation sites at the contact line) made in the model for droplets spreading on substrates are not applicable at low $\Delta T$. Conversely, for the complete range of $\Delta T$ we do not observe any preference of nucleation sites near the contact line. Modelling the contact line arrest at small undercooling requires to properly account for the statistical nature of the occurence of nucleations sites on the substrate, which is beyond the scope of this Letter.\\

\begin{figure}
 \centering
\includegraphics[width=1\linewidth]{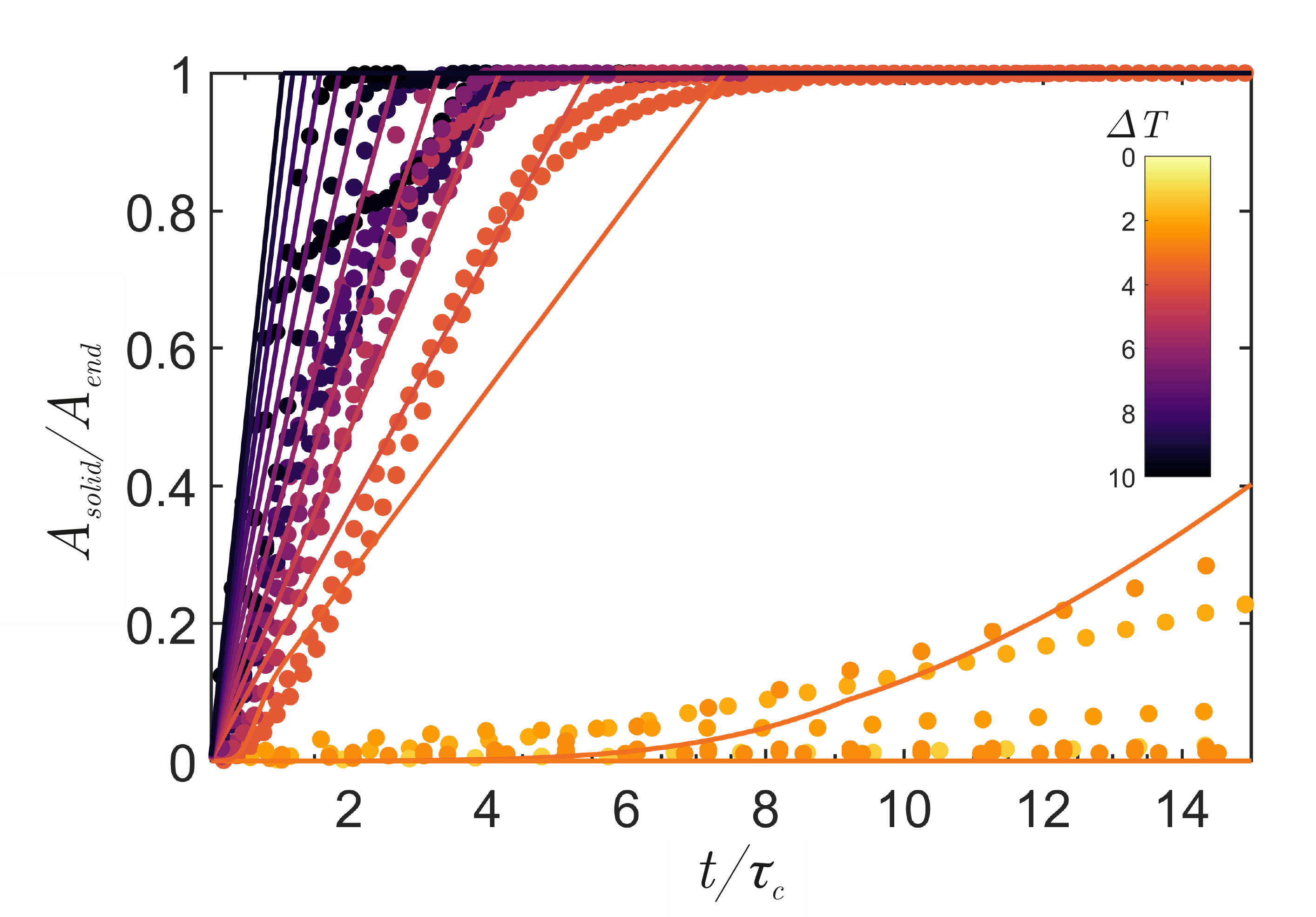}
\caption{Time dependence of the fraction of solidified material at the substrate for various temperatures. The solid lines show equation~(\ref{eq:JMAKscaled}), for $\SI{1}{\kelvin} < \Delta T < \SI{8}{\kelvin}$, see color code. A large change of solidification behaviour is seen for the model around $\Delta T = \SI{2.7}{\kelvin}$. Note that data with the same color can come from different experiments.}
\label{fig:solidification}
\end{figure}

Finally, we shift our focus to the temporal growth of the solidified area along the droplet-substrate interface. We follow the formulation proposed in Ref.~\cite{Zanotto1991, Fanfoni1998} for heterogeneous nucleation and growth of the solidified phase on an undercooled surface. It assumes time-independent growth velocity $U_{g}$ and nucleation rate $J_0$, with sites equally distributed over the substrate. We use the Johnson-Mehl-Avrami-Kolmogorov (\textit{JMAK}) equation to determine the 2D growth of the solidified surface fraction as a function of time as: $\chi=1-\exp\left(-4\pi N_0 U_{g}^{2}t^{2}\right)$. The amount of crystals per area is estimated as $N_0= \int_0^t J_0 \delta_{th} dt=2/3 J_0 \delta_{th} t$. As the droplet continues to spread over the substrate, the area available for nucleation increases. Hence, we rescale the \textit{JMAK} equation by the instantaneous wetted area $\left(R(t)\right)^2/R_{end}^2$, to find:
\begin{equation} \label{eq:JMAKscaled}
\frac{A_{\rm solid}}{A_{\rm end}} = \left(1-\exp\left(-\frac{8\pi}{3} J_0 \delta_{th} U_{g}^{2}t^{3}\right)\right) \left(\frac{R\left(t\right) }{R_{\rm end}}\right)^2,
\end{equation}
where we use the arrest criterion (equation \eqref{eq:clinearrest}) to obtain $R_{\rm end}$.\\

The evolution of the solidified area fraction for various surface undercoolings is shown in figure~\ref{fig:solidification}. For $\Delta T < \SI{2.5}~\rm K$, the slow nucleation leads to a very slow increase in solidified area fraction. At higher $\Delta T $, both the nucleation rate and crystal growth speed increase. Consequently, the solidified area fraction grows faster.  For even higher under-cooling $\Delta T > \SI{4.4}{\kelvin}$,  the solidified fraction growth rate matches the spreading of the droplet, thus: $A_{\mathrm{solid}}/A_{end} \approx \left(R(t)/R_{end}\right)^2$.   We find that equation~\eqref{eq:JMAKscaled} agrees with the experimental data over a wide range of $ \Delta T $. The only adjustable parameter is the geometrical factor $f \left( \theta_{ls} \right)$, since the average nucleus contact angle is not directly measurable~\cite{Maattanen2014}. From a fit to the data, we find $f(\theta_{ls})\approx 0.12$, which implies an average nucleus contact angle of $\theta_{ls}\approx \SI{55}{\degree}$. It must be pointed out that for undercooling below $2.5$ K, the model does not predict any solidification within the typical timescale of experiments. This observation corroborates our conjecture that it is impurities that cause nucleation and growth in this temperature range. \\

In summary, we directly visualized the surface solidification during spreading of hexadecane droplets on an under-cooled sapphire surface, using high-speed TIR imaging. Two distinct solidification behaviours are observed, which are explained by classical nucleation theory. The number of crystals in a spreading droplet scale with $n_{\rm{crystals}} \propto \left(t/\tau_c\right)^{5/2}$.
Furthermore, we reveal that the arrest velocity is approximately equal to the crystal growth velocity determined by the undercooling. This direct observation is in line with the model developed in Ref.~\cite{Ruiter2017}. However, it is not valid for very weak under-cooling, since the nucleation happens randomly over the surface. Apart from the processes near the contact line, we reveal that the crystal growth speed directly depends on the contact temperature $T_c$, rather than on the initial temperature of the substrate $T_s$.  
Finally, we showed that the 2-dimensional \textit{JMAK} equation, rescaled for the time dependent contact area accurately predicts the temporal growth of the solidified area fraction of a spreading droplet. 
Our results give insight into both local and overall solidification processes near the contact line and the substrate. The visualization method opens a new experimental pathway of elucidating solidification behavior near substrates on a time-resolved macro-and microscopic scale. This technique can be used to directly measure the early solidification behaviour for many relevant applications in manufacturing.

\section*{Acknowledgements}
We acknowledge funding by the Max Planck Center Twente, RBJK acknowledges funding by the TNO ERP programme 3D nanomanufacturing. KH acknowledges funding by German Science Foundation DFG within grant HA8467/1-1. DL acknowledges the ERC Advanced Grant DDD 740479. We thank Martijn van der Ouderaa and Sofie K\"olling for participating in the experiments.

\bibliography{MyCollection3}
\end{document}